\documentclass{emulateapj}

\newcommand{\etal}{et~al.}
\newcommand{\eg}{e.g., }
\newcommand{\ie}{i.e., }
\newcommand{\atel}{Astron.~Tel.}
\newcommand{\cbet}{Cent.~Bur.~Electron.~Tel.}
\newcommand{\Msun}{M_{\odot}}
\newcommand{\kms}{km~s$^{-1}$}
\newcommand{\ergs}{ergs~s$^{-1}$}

\newcommand{\Cofs}{$^{56}$Co}
\newcommand{\Nifs}{$^{56}$Ni}
\newcommand{\Mej}{M_{\rm ej}}

\newcommand{\Mni}{M{\rm (^{56}Ni)}}
\newcommand{\Mms}{M_{\rm ms}}
\newcommand{\Mc}{M_{\rm cut}}
\newcommand{\Mrem}{M_{\rm rem}}
\newcommand{\Mpre}{M_{\rm preSN}}
\newcommand{\Teff}{T_{\rm eff}}
\def\gsim{\mathrel{\rlap{\lower 4pt \hbox{\hskip 1pt $\sim$}}\raise 1pt
\hbox {$>$}}}
\def\lsim{\mathrel{\rlap{\lower 4pt \hbox{\hskip 1pt $\sim$}}\raise 1pt
\hbox {$<$}}}

\begin{document}

\title{The Peculiar Type Ib Supernova 2006jc: A
WCO Wolf-Rayet Star Explosion}

\author{
 N.~Tominaga\altaffilmark{1,2},
 M.~Limongi\altaffilmark{3,2,4},
 T.~Suzuki\altaffilmark{2}, 
 M.~Tanaka\altaffilmark{2},
 K.~Nomoto\altaffilmark{5,2},
 K.~Maeda\altaffilmark{5,6},
 A.~Chieffi\altaffilmark{7},
 A.~Tornambe\altaffilmark{3},
 T.~Minezaki\altaffilmark{8},
 Y.~Yoshii\altaffilmark{8},
 I.~Sakon\altaffilmark{2}, T.~Wada\altaffilmark{9},
 Y.~Ohyama\altaffilmark{9}, T.~Tanab\'e\altaffilmark{9},
 H.~Kaneda\altaffilmark{9}, T.~Onaka\altaffilmark{2},
 T.~Nozawa\altaffilmark{10}, T.~Kozasa\altaffilmark{10}, 
 K.~S.~Kawabata\altaffilmark{11},
 G.~C.~Anupama\altaffilmark{12}, 
 D.K.~Sahu\altaffilmark{12},
 U.K.~Gurugubelli\altaffilmark{12},
 T.P.~Prabhu\altaffilmark{12}, and
 J.~Deng\altaffilmark{13}
 }

\altaffiltext{1}{Optical and Infrared Astronomy Division, National
Astronomical Observatory, 2-21-1 Osawa, Mitaka, Tokyo, Japan;
nozomu.tominaga@nao.ac.jp}
\altaffiltext{2}{Department of Astronomy, School of Science,
University of Tokyo, 7-3-1 Hongo, Bunkyo-ku, Tokyo, Japan}
\altaffiltext{3}{Istituto Nazionale di Astrofisica - Osservatorio Astronomico di Roma, Via
Frascati 33, I-00040, Monteporzio Catone, Italy}
\altaffiltext{4}{Center for Stellar and Planetary Astrophysics, School
of Mathematical Sciences, P.O. Box, 28M, Monash University, Victoria 3800, Australia}
\altaffiltext{5}{Institute for the Physics and Mathematics of the Universe, 
University of Tokyo, Kashiwa, Chiba, Japan}
\altaffiltext{6}{Max-Planck-Institut f\"ur Astrophysik, Karl-Schwarzschild Strasse 1, 
85741 Garching, Germany}
\altaffiltext{7}{Istituto Nazionale di Astrofisica - Istituto di
Astrofisica Spaziale e Fisica Cosmica, Via Fosso del Cavaliere 100, Roma, Italy}
\altaffiltext{8}{Institute of Astronomy, School of Science, University
of Tokyo, 2-21-1 Osawa, Mitaka, Tokyo, Japan}
\altaffiltext{9}{Institute of Space and Astronautical Science, Japan
Aerospace Exploration Agency, 3-1-1 Yoshinodai, Sagamihara, Kanagawa, Japan}
\altaffiltext{10}{Department of Cosmosciences, Graduate School of Science,
Hokkaido University, Sapporo, Japan}
\altaffiltext{11}{Hiroshima Astrophysical Science Center, Hiroshima University, Hiroshima, Japan}
\altaffiltext{12}{Indian Institute of Astrophysics, Bangalore, 560 034, India}
\altaffiltext{13}{National Astronomical Observatories, Chinese Academy
of Sciences, Beijing 100012, China}

\setcounter{footnote}{13}

\begin{abstract}
 We present a theoretical model for Type Ib supernova (SN)~2006jc.
 We calculate the evolution of the progenitor star, hydrodynamics and 
 nucleosynthesis of the SN explosion, and the SN bolometric light curve
 (LC). The synthetic bolometric LC is compared with
 the observed bolometric LC constructed by
 integrating the UV, optical, near-infrared (NIR), and mid-infrared
 (MIR) fluxes.
 The progenitor is assumed to be as massive as $40\Msun$ on the zero-age
 main-sequence.
 The star undergoes extensive mass loss to reduce its mass down to as
 small as $6.9\Msun$, thus becoming a WCO Wolf-Rayet star.
 The WCO star model has a thick
 carbon-rich layer, in which amorphous carbon grains can be formed. This
 could explain the NIR brightening and the
 dust feature seen in the MIR spectrum.
 We suggest that the progenitor of SN~2006jc is a WCO Wolf-Rayet star
 having undergone strong mass loss and 
 such massive stars are the
 important sites of dust formation.
 We derive the parameters of the explosion model in
 order to reproduce the bolometric LC of SN~2006jc by the radioactive decays: 
 the ejecta mass $4.9\Msun$, hypernova-like explosion energy 
 $10^{52}$ ergs, and ejected \Nifs\ mass $0.22\Msun$. 
 We also calculate the circumstellar interaction and find that a CSM
 with a flat density structure is required to reproduce the X-ray LC of
 SN~2006jc. This suggests a drastic
 change of the mass-loss rate and/or the wind velocity that is
 consistent with the past luminous blue variable (LBV)-like event.
\end{abstract}


\section{INTRODUCTION}
\label{sec:introduction}

On 9th October 2006, \cite{nak06} reported K. Itagaki's discovery of a
possible supernova (SN) in UGC~4904. Although the SN was discovered
after the peak, an upper limit of the $R$ magnitude ($M_R>-12.2$) was obtained at
$\sim$20 days before the discovery \citep{pas07}. Interestingly,
\cite{nak06} also reported that an optical transient had appeared in
2004 close to the position of SN~2006jc. The transient was as
faint as $M_{\rm R}\sim-14$ and its duration was as short as $\sim10$
days. Since the event was faint and short-lived, they speculated that
the transient was a luminous blue variable (LBV)-like event. 
The spatial coincidence between the LBV-like event and SN~2006jc is
confirmed by \cite{pas07}.
Because of such an intriguing association with the LBV-like event,
many groups performed follow-up observations of SN~2006jc in various
wavebands: X-ray, ultra violet (UV), optical, infrared (IR), and radio.

Spectroscopic observations showed many broad features and strong narrow
\ion{He}{1} emission lines. According to the He detection, SN~2006jc
was classified as Type Ib 
\citep{cro06,fes06a,ben06,mod06a,mod06b}. However, strange spectral features and
their evolutions were reported. A bright blue continuum was prominent in
the optical spectrum at early epochs \citep{fol07,pas07,smi07}. Such a
bright blue continuum had also been observed in Type II SN~1988Z
\citep{tur93}, but the origin of this feature is still unclear. As
the blue continuum declined, the red wing brightened
and the optical spectra showed ``U''-like shapes \citep{smi07,kaw07b}.
This is a distinguishing feature of SN~2006jc in contrast to the spectra of
usual SNe that have a peak in optical bands.

Photometric observations in optical and IR bands were
performed continuously.
The optical light curve (LC) showed a rapid decline from 50 days after the
discovery, as in the case of SN~1999cq \citep{mat00}. 
At the same epoch, near infrared
(NIR) emissions brightened \citep{ark06,dic07}. The NIR brightness increased
from $\sim40$ days to $\sim70$ days after the discovery and
then declined \citep{dic07}. The epoch of the NIR brightening
corresponds to that of the development of the red wing in the optical
spectra \citep{smi07}. 

The NIR brightening, as well as the fact that the redder side of the He
emission profile declined faster than the bluer side, has been
interpreted as an evidence of an ongoing dust formation \citep{smi07}. Additionally, on 29th April
2007 (200 days after the discovery), the {\sl AKARI} satellite performed
NIR and mid-infrared (MIR) photometric and spectroscopic
observations \citep{sak07} and the {\sl MAGNUM} telescope obtained the
NIR photometries \citep{min07}. They report the formation of
amorphous carbon dust: another piece of evidences of the dust formation.

\begin{figure}
\epsscale{1.}
\plotone{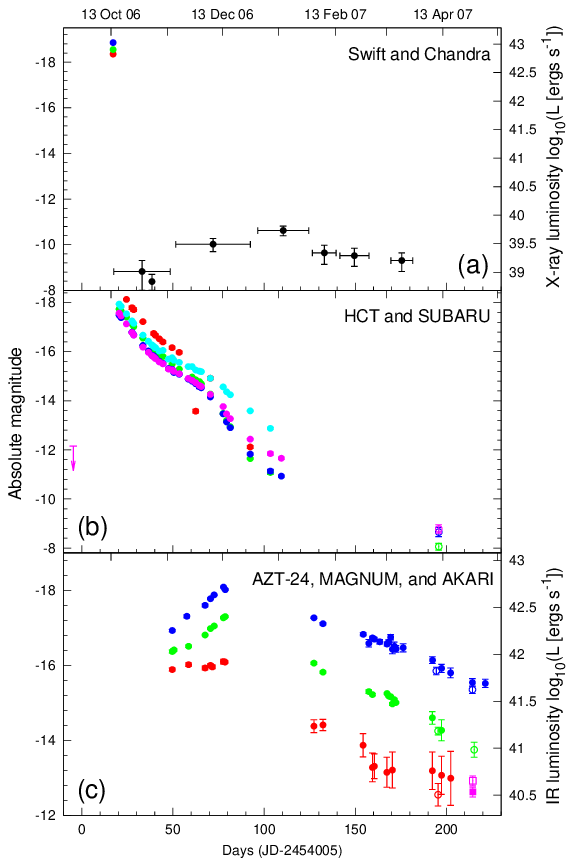}
\caption{Multicolor available observations of SN~2006jc. (a) X-ray and
 UV luminosities obtained with the {\sl Swift} and {\sl Chandra}
 satellites ({\it black}: X-ray, {\it red}: UVW2-band, {\it green}: UVM2-band, {\it
 blue}: UVW1-band, \citealt{imm07}). X-ray LC is shown in a unit of
 \ergs\ (right axis).
 (b) Optical luminosities obtained with the {\sl HCT} telescope ({\it
 filled}, \citealt{anu07})
 and the {\sl SUBARU} telescope ({\it open}, \citealt{kaw07a,kaw07b}). The upper limit is taken
 from \cite{pas07}. The color represents the
 wavelengths (U-band: {\it red}, B-band: {\it green}, V-band: {\it blue},
 R-band: {\it magenta}, I-band: {\it cyan}).
 (c) IR luminosities obtained with the {\sl AZT-24} telescope ({\it filled
 circles}, \citealt{ark06,dic07}), the {\sl MAGNUM} telescope ({\it open
 circles}, \citealt{min07}), and the
 {\sl AKARI} satellite ({\it squares}, \citealt{sak07}). The color of circles represents
 the wavelengths (J-band: {\it red}, H-band: {\it green}, and K-band: {\it blue}). The
 contributions to the IR luminosities from the hot dust ({\it filled square}) and with the hot
 and warm dust ({\it open square}) are shown in a unit of
 \ergs\ (right axis).
\label{fig:LCobsall}}
\end{figure}

X-ray and UV emissions have also been observed by the {\sl Swift} and {\sl
Chandra} satellites \citep{bro06,imm06,imm07,hol07}. X-ray observations were performed at seven
epochs and showed a brightening from $\sim20$ days to $\sim100$ days
after the discovery
\citep{bro06,imm06,imm07}. The X-ray detection suggests an interaction
between the SN ejecta and the circumstellar matter (CSM). On the
contrary, the radio emission was not detected by Very Large
Array (VLA) \citep{sod06}.

We present a SN explosion model of a Wolf-Rayet star that
explains the bolometric and X-ray LCs. Hydrodynamics, nucleosynthesis,
and LC synthesis calculations are performed assuming the spherical
symmetry. In this study, we assume the explosion date of
SN~2006jc to be 15 days before the discovery ($t=0$) and the energy source of
the light to be the \Nifs-\Cofs\ decay.

The paper is organized as follows: in \S~\ref{sec:bol}, we describe how we derive
the bolometric LC from observations in the various
wavebands, in \S~\ref{sec:preSN}, we briefly discuss the
presupernova evolutionary properties of the progenitor star; in
\S~\ref{sec:hyd}, hydrodynamical and nucleosynthesis calculations
are described; in \S~\ref{sec:LC}, LC synthesis calculations are
presented; in \S~\ref{sec:CSM}, we calculate the 
X-ray emission due to the ejecta-CSM interaction; in
\S~\ref{sec:conclude} and \S~\ref{sec:discuss}, conclusions and
discussion are presented.

\section{Photometric Observations and Bolometric Light Curve}
\label{sec:bol}

\begin{deluxetable}{cc}
 \tablecaption{Optical luminosities.\label{tab:UVopt}}
\tabletypesize{\footnotesize}
 \tablewidth{0pt}
 \tablehead{
   \colhead{Date}
 & \colhead{$L_{\rm opt}$} \\
   \colhead{[JD$-$2454005]}
 & \colhead{[$10^{40}{\rm ergs~s^{-1}}$]}
 }
\startdata
   20 & 370 \\
   21 & 340 \\
   24 & 250 \\
   27 & 180 \\
   28 & 170 \\
   33 & 110 \\
   36 & 87 \\
   38 & 75 \\
   39 & 70 \\
   40 & 66 \\
   42 & 58 \\
   44 & 53 \\
   47 & 44 \\
   49 & 40 \\
   53 & 36 \\
   58 & 28 \\
   60 & 27 \\
   62 & 25 \\
   64 & 23 \\
   65 & 22 \\
   70 & 15 \\
   77 & 6.3 \\
   79 & 4.8 \\
   81 & 4.0 \\
   89 & 2.2 \\
   92 & 2.1 \\
  103 & 1.0 \\
  119 & 0.36 \\
  138 & 0.23 \\
  195 & 0.15 
\enddata
\end{deluxetable}

The bolometric luminosities of SNe are usually estimated from the
integration over the optical and NIR emission because the usual SNe
radiate dominantly in the optical and NIR bands (\eg \citealt{yos03,min08}). However, the 
spectra of SN~2006jc show the bright red and blue wings
\citep{smi07,kaw07b,anu07}, which implies that the emissions in 
UV and IR bands considerably contribute to the bolometric luminosity.

We construct the bolometric luminosity with the integration of the
UV, optical, and IR photometries that are obtained with the {\sl HCT} \citep{anu07}, {\sl
AZT-24} \citep{ark06,dic07}, {\sl MAGNUM} \citep{min07}, and {\sl SUBARU} telescopes
\citep{kaw07a,kaw07b} and the {\sl Swift} \citep{imm07} and {\sl AKARI} satellites
\citep{sak07}. Since the UV fluxes are available only at $t=17$
days \citep{imm07}, the UV luminosity is estimated from the optical
luminosity at the other epoch. Available observations are shown in
Figure~\ref{fig:LCobsall}. Details of optical observations will be
presented in the forthcoming papers (\eg \citealt{anu07,kaw07b}). 
We adopt a distance of 25.8Mpc corresponding to a distance modulus of
32.05 \citep{pas07} and a reddening
of $E(B-V)=0.05$ \citep{sch98,pas07}.

\subsection{Optical emission}
\label{sec:OPTest}

\begin{figure}
\epsscale{1.}
\plotone{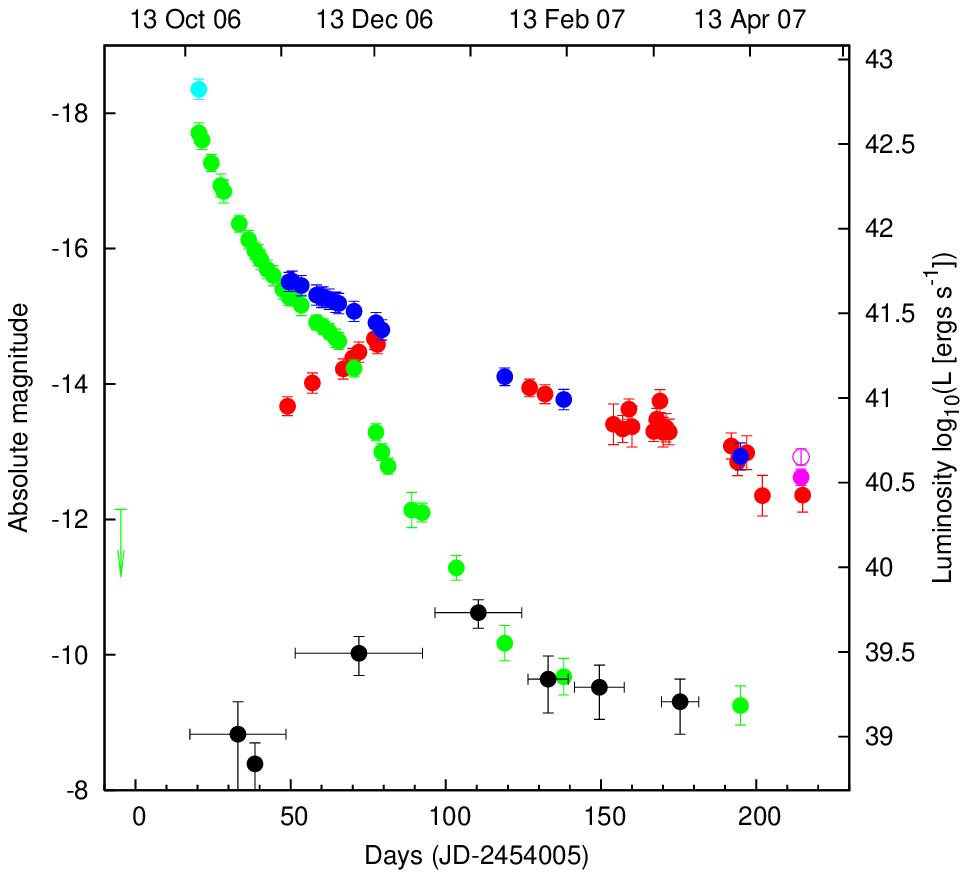}
\caption{Comparison of multicolor LCs of
 SN~2006jc ($L_{\rm opt}$: {\it green filled circles},
 \citealt{kaw07a,kaw07b,anu07}; $L_{\rm UV}+L_{\rm opt}$: {\it cyan
 filled circles}, \citealt{imm07,anu07}; $L_{\rm IR,est}(\nu<3\times10^{14}{\rm Hz})$: {\it red filled
 circles}, \citealt{ark06,dic07}; $L_{\rm opt}+L_{\rm IR,est}(\nu<3\times10^{14}{\rm Hz})$: {\it blue filled
 circles}, \citealt{kaw07a,kaw07b,anu07,ark06,dic07}; $L_{\rm IR,hot}(\nu<3\times10^{14}{\rm Hz})$:
 {\it magenta filled circle}, \citealt{sak07}; $L_{\rm IR}(\nu<3\times10^{14}{\rm Hz})$: {\it
 magenta open circle}, \citealt{sak07}; and $L_{\rm X}$: {\it black
 filled circles}, \citealt{imm07}).
\label{fig:LCobs}}
\end{figure}

The optical LCs were obtained with the {\sl HCT} and {\sl
SUBARU} telescopes \citep{kaw07a,kaw07b,anu07}. We integrate the optical
fluxes with a 
cubic spline interpolation from $3\times10^{14}$ Hz to
$1\times10^{15}$ Hz. The optical luminosities ($L_{\rm opt}$) 
are summarized in Table~\ref{tab:UVopt} and the LC is shown in
Figure~\ref{fig:LCobs}. The optical LC declines monotonically after the
discovery. The decline suddenly becomes rapid at $t>70$ days and the
optical luminosity finally goes down to 
$L_{\rm opt}\sim10^{39}$\ergs\ at $t\sim200$ days.

The X-ray LC obtained with the {\sl Swift} and 
{\sl Chandra} satellites \citep{imm07} shows that the X-ray luminosities,
$L_{\rm X}$, are much fainter than the optical luminosities
\citep{bro06,imm06,imm07}. Thus, the X-ray contribution to the
bolometric luminosities is negligible. However, the UV luminosity,
$L_{\rm UV}$, is comparable to the optical luminosity at $t=17$ days
($L_{\rm UV}\sim3\times10^{42}$\ergs\ as estimated from the
UVOT observations, \citealt{imm07}).\footnote{The UV flux is estimated
using the {\sl Swift} UVOT calibration documents
(http://swift.gsfc.nasa.gov/docs/swift/swiftsc.html).}
 The UV
luminosity is $\sim80\%$ of the optical luminosity, \ie the
total flux is $\sim1.8$ times brighter than
the optical flux (Fig.~\ref{fig:LCobs}). Since the UV flux
declined as the optical flux \citep{hol07}, we assume that
$L_{\rm UV}\sim0.8L_{\rm opt}$ at every epoch.
Although the blue wing declines with time and $L_{\rm UV}$ might be
over-estimated at $t\gsim90$ days \citep{smi07}, 
the bolometric luminosity ($L_{\rm bol}$) should be reliable because the IR
contribution dominates in the bolometric luminosity at such late epochs
(\S~\ref{sec:IRest}).

\subsection{Infrared emission}
\label{sec:IRest}

\begin{deluxetable*}{ccccc}
 \tablecaption{Parameters for amorphous carbon fitting of the
 JHK-band photometries and the estimated IR luminosities.\label{tab:NIRBB}}
\tabletypesize{\small}
 \tablewidth{0pt}
 \tablehead{
   \colhead{Date}
 & \colhead{$C_\epsilon$}
 & \colhead{$T_{\rm C,hot}$}
 & \colhead{$L_{{\rm IR,est}}(\nu<1.3\times10^{14}{\rm Hz})$}
 & \colhead{$L_{{\rm IR,est}}(\nu<3\times10^{14}{\rm Hz})$} \\
   \colhead{[JD$-$2454005]}
 & \colhead{[$10^{34}$]}
 & \colhead{[K]}
 & \colhead{[$10^{40}{\rm ergs~s^{-1}}$]}
 & \colhead{[$10^{40}{\rm ergs~s^{-1}}$]}
 }
\startdata
  49 & 3.9 & 1580 & 2.9 & 9.0 \\
  57 & 12 & 1330 & 5.1 & 12 \\
  67 & 16 & 1340 & 6.6 & 15 \\
  70 & 19 & 1330 & 7.8 & 17 \\
  72 & 23 & 1300 & 8.9 & 19 \\
  77 & 27 & 1310 & 11 & 23 \\
  79 & 18 & 1400 & 9.1 & 21 \\
 127 & 46 & 1050 & 7.6 & 12 \\
 132 & 52 & 1010 & 7.2 & 11 \\
 154 & 17 & 1150 & 4.2 & 7.0 \\
 157 & 32 & 1010 & 4.5 & 6.6 \\
 159 & 75 &  900 & 6.7 & 8.7 \\
 160 & 26 & 1050 & 4.4 & 6.8 \\
 167 & 35 &  990 & 4.5 & 6.4 \\
 168 & 54 &  940 & 5.5 & 7.5 \\
 169 & 99 &  880 & 7.6 & 9.7 \\
 170 & 48 &  930 & 4.8 & 6.3 \\
 171 & 45 &  950 & 4.9 & 6.7 \\
 172 & 44 &  940 & 4.6 & 6.3 \\
 192 & 48 &  900 & 4.0 & 5.3 \\
 195 & 45 &  870 & 3.3 & 4.2 \\
 197 & 56 &  860 & 3.8 & 4.8 \\
 202 & 5.6 & 1190 & 1.5 & 2.7 \\
 215 & 28 &  870 & 2.1 & 2.7 
\enddata
\end{deluxetable*}

The IR spectroscopy and photometries are obtained with the {\sl AZT-24}
and {\sl MAGNUM} telescopes (NIR photometries, \citealt{ark06,dic07,min07})
and the {\sl AKARI} satellite (NIR spectroscopy and MIR photometries,
\citealt{sak07}). As indicated by the red wing in the optical spectra, the IR
emission considerably contributes to the bolometric luminosity of
SN~2006jc.

The MIR observation is available at $t=215$
days \citep{sak07}. The IR
luminosity integrated over $\nu<3\times10^{14}$ Hz
is estimated from the NIR and MIR observations as  
$L_{{\rm IR}}(\nu<3\times10^{14}{\rm Hz})=4.5\times10^{40}$ \ergs. 
\cite{sak07} concluded that the IR emission is originated from
amorphous carbon grains with two temperatures of $T=800$K and 320K. 
The large difference between the two temperatures would
imply that the origin of the hot carbon dust with $T=800$K is different
from that of the warm carbon dust with $T=320$K. The hot carbon dust is
suggested to be newly formed in the SN ejecta and heated by the
\Nifs-\Cofs\ decay by a dust formation calculation \citep{noz07}.
On the other hand, the origin of the emission from the warm carbon dust is
suggested to be a SN light echo of the CSM carbon dust
(\citealt{sak07,mat08}; see also \citealt{noz07}).
Therefore, we assume that the optical emission from SN~2006jc is
absorbed and simultaneously re-emitted by the hot carbon dust and thus
the luminosity emitted from the hot carbon dust should be included
in the bolometric luminosity of SN~2006jc.
According to the estimated temperatures and masses of the hot and
warm carbon grains \citep{sak07}, the luminosities contributed by the hot and warm
carbon grains are 
$L_{{\rm IR,hot}}(\nu<3\times10^{14}{\rm Hz})=3.2\times10^{40}$ \ergs\ and 
$L_{{\rm IR,warm}}(\nu<3\times10^{14}{\rm Hz})=1.1\times10^{40}$ \ergs,
respectively.\footnote{The difference between 
$L_{{\rm IR}}(\nu<3\times10^{14}{\rm Hz})$ and 
$L_{{\rm IR,hot}}(\nu<3\times10^{14}{\rm Hz})+L_{{\rm IR,warm}}(\nu<3\times10^{14}{\rm Hz})$ 
stems from that the H-band luminosity is slightly brighter than the
luminosity emitted from the hot carbon dust \citep{sak07}.}

For the epochs when the IR photometries at $\nu<1.3\times10^{14}{\rm Hz}$
are unavailable, we estimate the contribution of the IR emission by
fitting the JHK-band photometries with amorphous carbon emission.

From the Kirchhoff's law, the thermal radiation from
a spherical dust grain X with a uniform radius $a_{\rm X}$ and
temperature $T_{\rm X}$ is given by 
$4\pi a_{\rm X}^2 B(\nu,T_{\rm X})Q^{\rm abs}_{\rm X}(\nu)$,
where $Q^{\rm abs}_{\rm X}(\nu)$ is the absorption efficiency of the
grain. For the optically thin case, the observed emission from dust
grains $X$ is written as
\begin{equation}
 f_{\rm X}(\nu)=N_{\rm X} \pi B(\nu,T_{\rm X})Q^{\rm abs}_{\rm
		  X}(\nu) \left({a_{\rm X}\over{R}}\right)^2,
\end{equation}
where $N_{\rm X}$ and $R$ denote the total number of the dust particles
and the distance from the observer, respectively \citep{sak07}. 
In the followings, we convolve the
$\nu$-independent coefficients as an emission coefficient 
$C_\epsilon=\pi N_{\rm X} \left({a_{\rm X}/R}\right)^2$. 
Applying the absorption efficiency for the amorphous carbon grain with
$a_{\rm C}=0.01{\rm \mu m}$, we derive the temperature of the hot carbon
dust, $T_{\rm C,hot}$, and $C_\epsilon$ to reproduce the JHK-band
photometries.

To justify the above estimate, we compare the estimate with the actual MIR
observation at $t=215$ days \citep{sak07}. The fitting gives the temperature
$T_{\rm C,hot}=870$ K and the emission coefficient $C_\epsilon=2.8\times10^{35}$
for the HK-band photometries at $t=215$ days. The luminosity
integrated over $\nu<3\times10^{14}{\rm Hz}$ is 
$L_{{\rm IR,est}}(\nu<3\times10^{14}{\rm Hz})=2.7\times10^{40}$
\ergs. The temperature and luminosity are roughly consistent with those
of the hot carbon dust. The agreement indicates that the 
fitting gives a good estimate of the IR emission due to the hot carbon
dust. We note that the estimate can not account for the emission from the
warm carbon dust.

Table~\ref{tab:NIRBB} summarizes the emission coefficient, 
temperature, estimated luminosity at $\nu<1.3\times10^{14}$ Hz,
and luminosity emitted below $\nu=3\times10^{14}$ Hz. 
The dust temperature roughly declines from $T_{\rm C,hot}\sim1600$K at
$t=49$ days to $T_{\rm C,hot}\sim870$K at $t=215$ days. This is consistent with a
picture that the hot carbon dust was formed in the SN ejecta and cooled down
gradually \citep{noz07}. The IR LC is shown in Figure~\ref{fig:LCobs}. The
estimated luminosity at $\nu<1.3\times10^{14}$ Hz evolves as the JHK
LCs, and thus the IR LC brightens at $t\sim50-80$ days and declines
at $t>120$ days. Since there is no NIR data at 
$t\sim80-120$ days, the bolometric LC can not be estimated at this
epoch. The bolometric luminosity is derived from the summation of 
$L_{\rm UV}$, $L_{\rm opt}$, and $L_{\rm IR}$ and summarized in
Table~\ref{tab:bol}, where $L_{\rm UV}=0.8 L_{\rm opt}$ is applied.

\begin{deluxetable}{cc}
 \tablecaption{Bolometric luminosities.\label{tab:bol}}
 \tablewidth{0pt}
 \tablehead{
   \colhead{Date}
 & \colhead{$L_{\rm bol}$}\\
   \colhead{[JD$-$2454005]}
 & \colhead{[$10^{40}{\rm ergs~s^{-1}}$]}
 }
 \startdata
   49 &  81 \\
   51 &  81 \\
   53 &  75 \\
   58 &  64 \\
   60 &  61 \\
   62 &  59 \\
   65 &  55 \\
   66 &  54 \\
   70 &  45 \\
   77 &  33 \\
   79 &  29 \\
  119 &  14 \\
  138 &  10 \\
  195 &  4.7
 \enddata
\end{deluxetable}

\section{The progenitor star}
\label{sec:preSN}

\begin{figure*}
\epsscale{.8}
\plotone{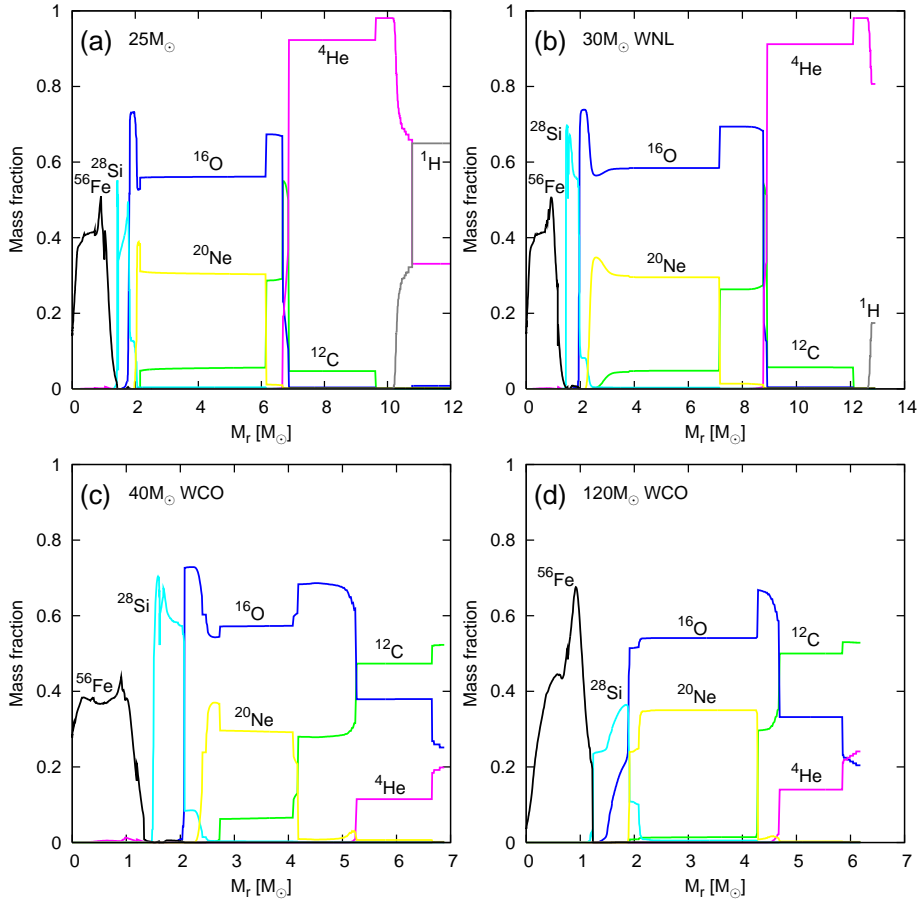}
\caption{Abundance distributions of presupernova models with (a)
 $\Mms=25\Msun$, (b) $\Mms=30\Msun$, (c) $\Mms=40\Msun$, and (d)
 $\Mms=120\Msun$. Note that the chemical composition of the outmost
 layer is C rich in the WCO Wolf-Rayet stars (c,d).
\label{fig:modpresn}}
\end{figure*}

The presupernova model has been extracted from a set of models already
presented by \cite{lc06} and computed with the latest release of
the stellar evolutionary code FRANEC (5.050218). 
Since all the features of this code have been already presented,
we will address here only the main points.
The interaction between convection and local nuclear burning
has been taken into account by coupling together and solving
simultaneously the set of equations
governing the chemical evolution due to the nuclear reactions and
those describing the convective mixing. More specifically, the
convective mixing has been treated by means of a diffusion equation
where the diffusion coefficient is computed by the use
of the mixing-length theory.
The nuclear network is the same as that adopted in \cite{lc03}, but
the nuclear cross sections have been updated whenever possible
(see Table 1 in \citealt{lc06}). A moderate amount of overshooting
of 0.2 $\rm H_{p}$ has been included into the calculation only on
the top of the convective core during core H burning.
Mass loss has been taken into account following the prescriptions of 
\cite{val00} for the blue supergiant phase ($\Teff>12000$K),
\cite{dejag88} for the red supergiant phase ($\Teff<12000$K),
\cite{nl00} for the WNL Wolf-Rayet phase and \cite{la89} during the
WNE/WCO Wolf-Rayet phases. We adopt the following correspondence of the
models to the various WR phases according to the surface abundances, as suggested
by \cite{mm03}: WNL 
($10^{-5}<X({\rm H})_{\rm surf}<0.4$), WNE ($X({\rm H})_{\rm surf}<10^{-5}$ and 
$\rm (C/N)_{\rm surf}<0.1$), WNC ($\rm 0.1<(C/N)_{\rm surf}<10$) and WCO 
($\rm (C/N)_{\rm surf}>10$). (Hereafter, C/N and C/O denote the number ratios.)

The X-ray emission, as well as the early bright blue continuum
and the narrow \ion{He}{1} lines, clearly indicates an interaction between the SN ejecta
and the CSM, \ie the existence of a dense CSM. Furthermore, the IR
spectral energy distribution may be explained by the formation of
amorphous carbon grains in the SN ejecta and the CSM (\citealt{sak07},
see also \citealt{noz07}). Since the C-rich environment (i.e., $\rm
C/O>1$) is required to form carbon dust (\eg
\citealt{noz03}), the IR observations suggest that the SN
ejecta and CSM contain a C-rich layer. This suggests that
the progenitor star of SN~2006jc is a WCO Wolf-Rayet star
with a C-rich envelope and CSM
(Figs.~\ref{fig:modpresn}a-\ref{fig:modpresn}d).

\begin{deluxetable}{lr}
\tablewidth{0pt}
\tablecolumns{2}
\tabletypesize{\footnotesize}
\tablecaption{Basic evolutionary properties of the progenitor star.\label{tab:presn}}
\tablehead{\multicolumn{2}{c}{Key quantities}}
\startdata 
\sidehead{\bf H Burning}
\cline{1-2} \\
$t_{\rm H}$ [Myr]                     &   4.64          \\
$M_{\rm CC}$ [$\Msun$]              &  25.80          \\
$M_{\rm tot}$ [$\Msun$]             &  35.40          \\
$t_{\rm O}$ [Myr]                     &   4.16          \\
$M_{\rm He}$ [$\Msun$]              &  10.01          \\
\sidehead{\bf He Burning}
\cline{1-2} \\
$t_{\rm He}$ [Myr]                    &   0.46          \\
$M_{\rm He,CC}$ [$\Msun$]              &  12.56          \\
$M_{\rm tot}$ [$\Msun$]             &   7.04          \\
$M_{\rm env}$ [$\Msun$]             &  18.80          \\
$X(\rm ^{12}C)_{\rm cen}$                        &   0.28          \\
$t_{\rm red}$ [Myr]                   &   0.07          \\
$t_{\rm WNL}$ [Myr] ($X{\rm (He)_{cen}}$)    &   0.11 (0.77)   \\ 
$t_{\rm WNE}$ [Myr] ($X{\rm (He)_{cen}}$)    &   0.054 (0.44)   \\
$t_{\rm WCO}$ [Myr] ($X{\rm (He)_{cen}}$)    &   0.21 (0.31)   \\
\sidehead{\bf Advanced Burnings}
\cline{1-2} \\
$\Delta t_{\rm exp}$ [yr]            &   1.25(+4)      \\ 
$M_{\rm He} {\rm (max)}$ [$\Msun$]        &  16.52          \\
$M_{\rm CO} {\rm (max)}$ [$\Msun$]        &   4.83          \\
$M_{\rm Fe,preSN}$ [$\Msun$]              &   1.50          \\ 
$\Mpre$ [$\Msun$]             &   6.88          \\ 
$R_{\rm preSN}$ [cm]                    &   3.08(+10)     \\
$M_r(\rm He_{shell})$ (\rm Int.-Ext.) [$\Msun$]         & 5.262-6.648     \\  
$M_r(\rm C_{shell})$ (\rm Int.-Ext.) [$\Msun$]          & 2.736-4.097     \\
$t_{\rm WNL} {\rm (tot)}$ [yr]              &   1.10(+5)      \\ 
$t_{\rm WNE} {\rm (tot)}$ [yr]              &   5.43(+4)      \\
$t_{\rm WCO} {\rm (tot)}$ [yr]               &   2.21(+5)      \\ 
$t_{\rm WR} {\rm (tot)}$ [yr]               &   3.86(+5)      
\enddata            
\end{deluxetable}

Inspection of all the presupernova models available in \cite{lc06}
indicates that only massive models, \ie $\Mms>40\Msun$, fulfill
the requirements from the IR observation and become WCO stars.
Moreover, these are the only stars in which
the chemical compositions of the mantle and CSM are dominated mainly by
C with a smaller amount of O (Fig.~\ref{fig:modpresn}cd).

In stars with initial masses smaller than $\Mms\sim 35\Msun$, the mass
of the He convective core increases or remains constant during the core He
burning phase. At core He exhaustion, a sharp discontinuity of He
abundance is produced at the outer edge of the CO core. Then, 
the CO core begins to contract to ignite the next nuclear fuel
while He burning shifts to a shell inducing a formation of a convective zone.
The He convective shell forms beyond the He 
discontinuity at the outer edge of the CO core. Hence 
its chemical composition is dominated by He [$X$(He)$>0.9$].
Because of the short lifetime of the
advanced burning stages, only a small amount of He is burned inside the
shell before the presupernova stage (Figs.~\ref{fig:modpresn}ab). 
Such a behavior is typical for stars in which the He core mass
remains roughly constant during core He burning (\eg \citealt{nom88}).

In stars with initial masses greater than $\Mms\sim 35\Msun$, on the contrary,
the mass loss is efficient enough ($10^{-5}-10^{-4}\Msun~{\rm yr}^{-1}$) to uncover
the He core and they reduce progressively their mass during the core He burning phase.
The star enters the WNE Wolf-Rayet stage and its subsequent evolution is
governed by the actual size of the He core. In particular, as the He
core progressively reduces due to the mass loss, the star tends to
behave as an initially-lower mass star, i.e., 
essentially reduces its central temperature. This induces the He
convective core to shrink progressively in mass as well, leaving a layer with
a variable chemical composition that reflects the central abundances
at various stages during core He burning.
When the stellar mass is reduced below the maximum extension
of the He convective core, the products of core He burning appear on the
surface and the star becomes a WCO Wolf-Rayet star. At core He
exhaustion, He burning shifts to a shell inducing the formation of
the convective shell. The convective shell forms in the region
with variable chemical composition. As a consequence,
at variance with what happens in stars with $\Mms\lsim35\Msun$,
in these stars, the chemical composition of the
convective shell becomes a mixture of the central He burning products.
Hence it is mainly composed of C, O and He (Figs.~\ref{fig:modpresn}cd).

Since all the models above $40~M_\odot$ have a similar presupernova structure,
we selected a $40~M_\odot$ star as representative of a typical star becoming a WCO 
Wolf-Rayet star. The mass at the presupernova stage ($\Mpre$) is
$\Mpre=6.9\Msun$ because of the strong mass loss. 
We underline that the $\Mms$-$\Mpre$ relation
is highly uncertain because it strongly depends on many details of the stellar
evolution (\eg the mass loss, overshooting, rotation, and metallicity, see
\citealt{la89,nl00,mey03,nom06,lc06,eld06}). 
For this reason, for the purpose of this study,
we mainly focus on a WCO progenitor with 
$\Mpre\sim 6.9$ $\rm M_\odot$, without paying much emphasis on 
$\Mms$.

\begin{figure}
\plotone{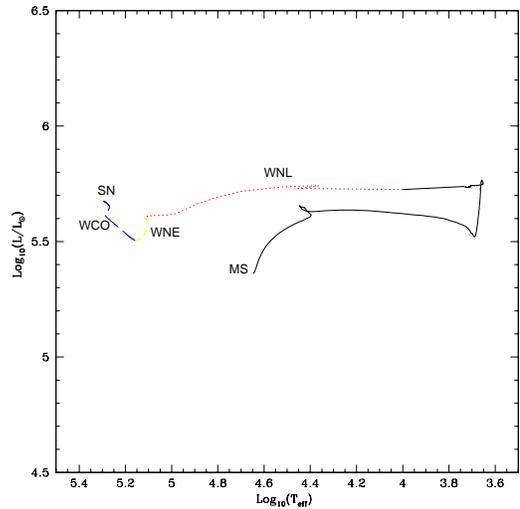}
\caption{Presupernova evolutionary path of the progenitor star with
 $\Mms=40\Msun$ in the Hertzsprung-Russell diagram.
\label{fig:hr40}}
\end{figure}

\begin{figure}
\plotone{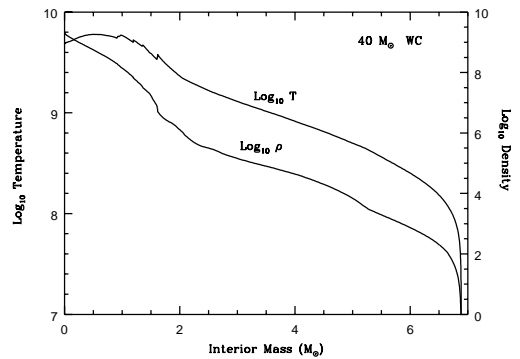}
\caption{Temperature and density structures of the presupernova
 progenitor star with $\Mms=40\Msun$.
\label{fig:40pre2}}
\end{figure}

A detailed discussion of the presupernova evolution during all
the nuclear burning stages is beyond the purpose of this paper. Hence
we report here in Table~\ref{tab:presn} some key properties during
the H, He, and advanced burning stages. In particular, 
for the H burning stage we report the following quantities:
the H burning lifetime ($t_{\rm H}$),
the maximum extension of the convective core ($M_{\rm CC}$), the total
mass ($M_{\rm tot}$) at core H exhaustion,
the time spent as an O-type star ($t_{\rm O}$) and the He core mass ($M_{\rm He}$) at H
exhaustion. Here, we assume that
the temperature of the O-type stars is $33000~{\rm K}<\Teff<50000~{\rm K}$.
For the He burning phase we report the following
quantities: the He burning lifetime ($t_{\rm He}$),
the maximum size of the He convective core ($M_{\rm He,CC}$), $M_{\rm tot}$ at core He exhaustion,
the maximum depth of the convective envelope ($M_{\rm env}$), the central
$\rm ^{12}C$ mass fraction at core He exhaustion [$X(\rm ^{12}C)_{\rm cen}$], 
the time spent at the red side ($\log \Teff<3.8$) of the HR diagram
($t_{\rm red}$), and the WNL, WNE, and WCO lifetimes ($t_{\rm WNL}$,
$t_{\rm WNE}$, and $t_{\rm WCO}$, respectively) - in parenthesis the
central He mass fraction [$X{\rm (He)_{cen}}$] when the star enters the
WNL, WNE and WCO phases.
For the advanced burning stage we report the following key quantities:
the time until the explosion ($\Delta t_{\rm exp}$), the maximum size of
the He core [$M_{\rm He} {\rm (max)}$], the maximum
size of the CO core [$M_{\rm CO} {\rm (max)}$], the masses of the
iron core ($M_{\rm Fe,preSN}$) and the star ($\Mpre$) and the radius of
the star ($R_{\rm preSN}$) at the presupernova stage, 
the final extension in mass of the He convective shell 
[$M_r({\rm He_{shell}})$] and of the convective C shell 
[$M_r({\rm C_{shell}})$], and
the total lifetimes during the WNL [$t_{\rm WNL} {\rm (tot)}$], WNE 
[$t_{\rm WNE} {\rm (tot)}$], WCO [$t_{\rm WCO} {\rm (tot)}$], and
WR [$t_{\rm WR} {\rm (tot)}$, where $t_{\rm WR}=t_{\rm WNL}+t_{\rm WNE}+t_{\rm WCO}$] phases.

Figures \ref{fig:hr40} and \ref{fig:40pre2} show the evolutionary path in the HR diagram
and the temperature and density profiles 
at the presupernova stage. 

\begin{figure}
\epsscale{1.}
\plotone{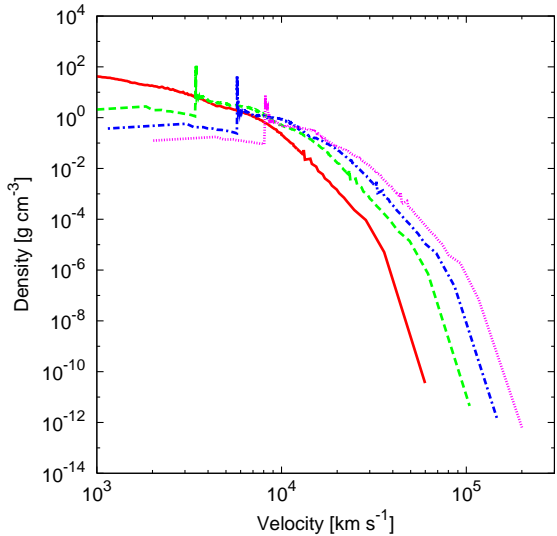}
\caption{Density structures at 100 s after the explosions for the models 
 with $E_{51}=1$ ({\it solid}), $E_{51}=5$ ({\it dashed}),
 $E_{51}=10$ ({\it dash-dotted}), and $E_{51}=20$ ({\it dotted}).
\label{fig:vrho}}
\end{figure}

\section{Hydrodynamics and Nucleosynthesis}
\label{sec:hyd}

The SN explosion and explosive nucleosynthesis are calculated for the
progenitor star with $\Mpre=6.9\Msun$. We apply various explosion
energies ($E_{51}=E/10^{51}~{\rm ergs}=1,~5,~10,~{\rm and}~20$) for the
SN explosion calculations (\eg \citealt{nom06,tom07c}). The hydrodynamical calculation is
performed by means of a spherical Lagrangian hydrodynamics code with a piecewise parabolic
method (PPM, \citealt{col84}) including nuclear energy production from
the $\alpha$-network. The equation of state takes account of the gas, radiation,
${\rm e}^-$-${\rm e}^+$ pair \citep{sug75}, Coulomb interactions between ions and electrons, and phase
transition \citep{nom82,nom88}. After the hydrodynamical calculations,
nucleosynthesis is calculated as a post-processing with a reaction
network that includes 280 isotopes up to $^{79}$Br (see Table~1 in
\citealt{ume05}).

Since the explosion mechanism of a core-collapse SN for
a massive star with an iron core is still an unsolved problem (\eg \citealt{jan07}), we
initiate the SN explosion as a thermal bomb. Although there are various
ways to simulate the explosion (\eg a kinetic piston, \citealt{woo95}),
it is suggested that the explosive nucleosynthesis does not depend
sensitively on the way how the explosion energy is deposited \citep{auf91}.
We set an inner reflective boundary at $M_r=1\Msun$ and $r=1000$~km
within the iron core and elevate temperatures at the inner boundary. 

In the spherical symmetry case, for any given progenitor model,
hydrodynamics and nucleosynthesis are determined by the explosion
energy. During the SN explosion, a shock propagates outward
inducing local compression and heating, triggering explosive
nucleosynthesis. Behind the shock front the matter is accelerated
and starts moving outward. However, if the progenitor has a deep gravitational
potential and the explosion energy is low, the inner layers begin to
fall back due to the gravitational attraction. A more
compact star and a lower explosion energy leads a larger amount of
fallback. The fallback has a deep implication on the SN nucleosynthesis
because it decreases the matter ejection, especially, of the
inner core (\eg \Nifs). 

\begin{figure}
\epsscale{1.}
\plotone{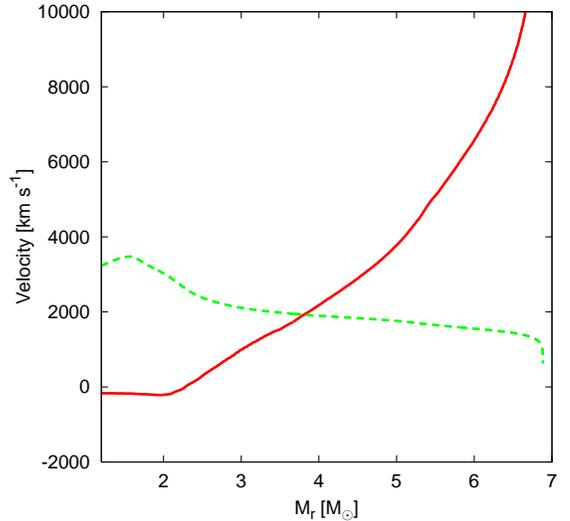}
\caption{Comparison between an escape velocity ({\it dashed}) and an
 ejecta velocity for the model with $E_{51}=1$ ({\it solid}).
\label{fig:escape}}
\end{figure}

Figure~\ref{fig:vrho} shows density structures at 100 s after the
explosions when homologously expanding structures are reached
($v\propto r$). We find that the fallback takes place for the model with
$E_{51}=1$ but not for the models with $E_{51}=5$, 10, and 20. 
Figure~\ref{fig:escape} shows a comparison between the escape velocity
and the ejecta velocity for the model with $E_{51}=1$ and demonstrates
that the matter below $M_r=M_{\rm fall}=3.8\Msun$ will fall back.
On the other hand, in the models with $E_{51}=5$, 10, and 20,
the matter above the inner boundary will be ejected. 

\begin{figure*}
\epsscale{.8}
\plotone{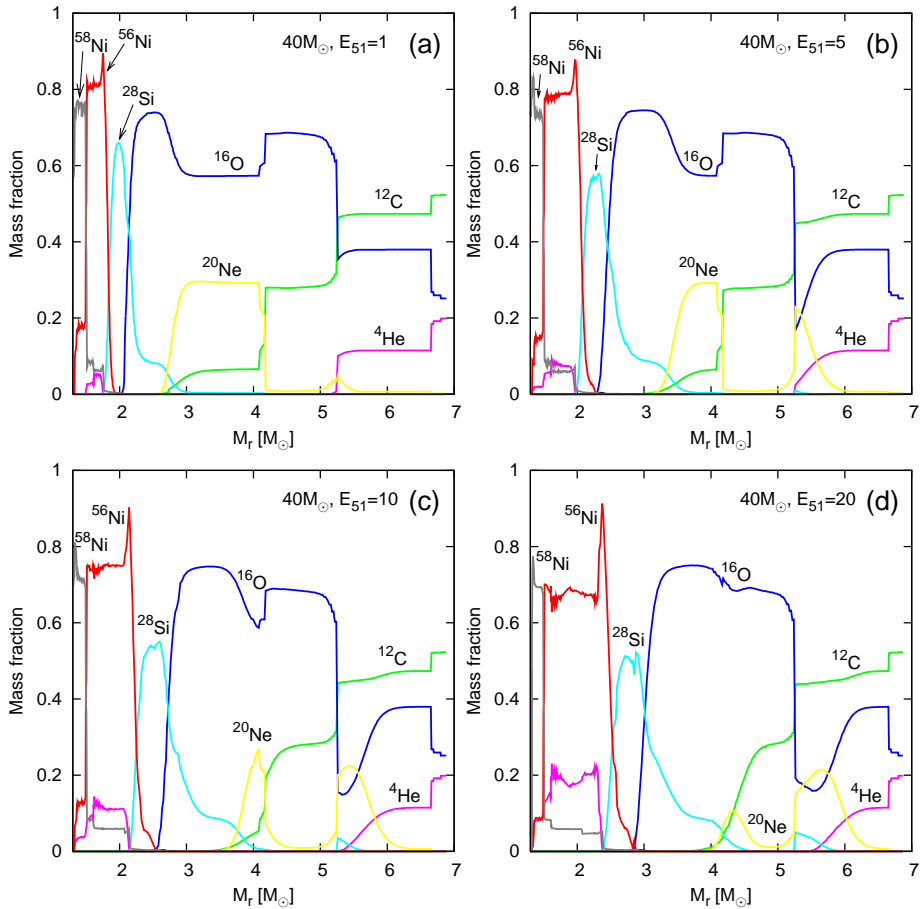}
\caption{Abundance distributions after the explosions of the progenitor star with
 $\Mms=40\Msun$. The explosion energies are (a) $E_{51}=1$,
 (b) $E_{51}=5$, (c) $E_{51}=10$, and (d) $E_{51}=20$.
\label{fig:abn}}
\end{figure*}

The abundance distributions after the explosions are shown in
Figures~\ref{fig:abn}a-\ref{fig:abn}d. In every model, $^{58}$Ni is
synthesized in the innermost layer ($M_r<M_{\rm Fe,preSN}=1.5\Msun$) due
to the low electron fraction.
Thus, we can estimate the maximum amounts of synthesized \Nifs\ for
given energies. The \Nifs-rich layer 
extending to $M_r=M_{\rm Fe}$ [where  $X({\rm ^{56}Ni})=X({\rm ^{28}Si})$] and
$^{28}$Si-rich layer extending to $M_r=M_{\rm Si}$ [where 
$X({\rm ^{28}Si})=X({\rm ^{16}O})$] expand farther in the models with
higher $E$ because the temperature achieved is higher in the outer layer
for higher $E$. $M_{\rm Fe}$ and $M_{\rm Si}$ for each model are
summarized in Table~\ref{tab:abn}.

\begin{deluxetable}{ccccc}
 \tablecaption{Nucleosynthesis properties of the explosion models with
 $\Mpre=6.9\Msun$ ($\Mms=40\Msun$). \label{tab:abn}}
 \tablewidth{0pt}
 \tablehead{
   \colhead{Explosion energy}
 & \colhead{$M_{\rm Fe}$} 
 & \colhead{$M_{\rm Si}$} 
 & \colhead{$\Mni$} 
 & \colhead{$\Mc$} \\
   \colhead{[$10^{51}$ ergs]}
 & \colhead{[$\Msun$]}
 & \colhead{[$\Msun$]}
 & \colhead{[$\Msun$]}
 & \colhead{[$\Msun$]}
 }
\startdata
 1 & 1.8 & 2.1 & --- & --- \\
 5 & 2.1 & 2.5 & 0.5 & 1.8 \\
 10 & 2.3 & 2.7 & 0.6 & 2.0 \\
 20 & 2.5 & 3.0 & 0.7 & 2.3 
\enddata
\end{deluxetable}

\Nifs\ is synthesized at $M_r<M_{\rm Si}$. Since 
$M_{\rm Si}<M_{\rm fall}$ in the model with $E_{51}=1$, the model is
likely not to eject \Nifs. 
On the other hand, the models with $E_{51}=5$, 10, and 20 can eject all
synthesized \Nifs\ because the fallback does not occur. The total
amounts of synthesized \Nifs\ for the models with $E_{51}=5$, 10, and 20
are summarized in Table~\ref{tab:abn}. 

\vspace{1cm}

\section{Light Curve}
\label{sec:LC}

The energy source of the LC of SN~2006jc is still under debate. The
possible sources include the \Nifs-\Cofs\ decay like Type I SNe and the ejecta-CSM
interaction like Type IIn SNe. However, both scenarios have the following problems. 
In the case of the \Nifs-\Cofs\ decay, the $\gamma$-ray photon and positron
emitted from the \Nifs-\Cofs\ decay are absorbed by the SN ejecta and
the absorbed energy is thermalized. Thus, the spectra would show a
blackbody-like continuum as normal Type I SNe do. However, the spectra
of SN~2006jc do not resemble those of normal Type I SNe but show a
bright blue continuum in early epochs \citep{fol07,smi07}. In the case of the ejecta-CSM
interaction, the kinetic energy is transformed to an X-ray emission via
bremsstrahlung radiation, and then converted to UV, optical, and IR
emissions. Thus, it is difficult to explain that the X-ray luminosity is
much fainter than the optical luminosity unless the optical depth for
the X-ray emission is much higher than that for the optical
emission. Another problem with the ejecta-CSM interaction model is that
the X-ray LC is not synchronized with the bolometric LC. In addition,
the LC powered by the ejecta-CSM interaction usually has a long-term
plateau (\eg SN~1997cy, \citealt{tur00}). Thus, we assume that the LC is
powered by the \Nifs-\Cofs\ decay.

\vspace{1cm}

\subsection{Radioactive Decay Models} 
\label{sec:56ni}

The bolometric LC of SN~2006jc is constructed from the UV, optical, and IR
observations as described in \S~\ref{sec:bol}. The estimated peak
bolometric magnitude of SN~2006jc is $M=-18.4$, being as bright as
SN~2006aj (\eg \citealt{pia06}). Thus, it is speculated that the ejected
amount of \Nifs\ [$\Mni$] is similar to SN~2006aj, \ie
$\Mni\sim0.2\Msun$ \citep{maz06,maz07,mae07}.  According to
\S~\ref{sec:hyd}, the models with $E_{51}=5$, 10, and 20 can eject a
large enough amount of \Nifs, while the \Nifs\ production of the model with
$E_{51}=1$ is too small.

The spherical explosion models with $E_{51}=5$, 10, and 20 yield too
much $\Mni$ because of no fallback. However, no fallback is
a consequence of the assumption of the spherical symmetry. 
The fallback takes place in an aspherical explosion even
with a high explosion energy and thus the aspherical explosion may
well decrease $\Mni$ and increase the central remnant mass
$\Mrem$ (\citealt{mae03,tom07a,tom07b}). Therefore, assuming that 
aspherical fallback takes place in the high-energy models with 
$E_{51}=5$, 10, and 20, we estimate the amount of fallback to
yield $\Mni\sim0.2\Msun$ and then the ejected masses for the models as
$\Mej=\Mpre-\Mrem$. As a result, the sets of $\Mrem$, $\Mej$ and $E$
are derived to be ($\Mrem/\Msun$, $\Mej/\Msun$, $E_{51}$) $=$ (1.8, 5.1,
5), (2.0, 4.9, 10), and (2.3, 4.6, 20).

Applying the homologous density structures of the models (Fig.~\ref{fig:vrho}), 
we synthesize bolometric LCs for the models with $E_{51}=5$, 10, and 20
using the LTE radiation hydrodynamics code and the gray $\gamma$-ray transfer 
code \citep{iwa00}. In the radiative transfer calculation, the electron
scattering is calculated for the ionization states solved by the saha equation
and the Rosseland mean opacity is approximated with an empirical
relation to the electron-scattering opacity \citep{den05}. 

\begin{figure}
\epsscale{1.}
\plotone{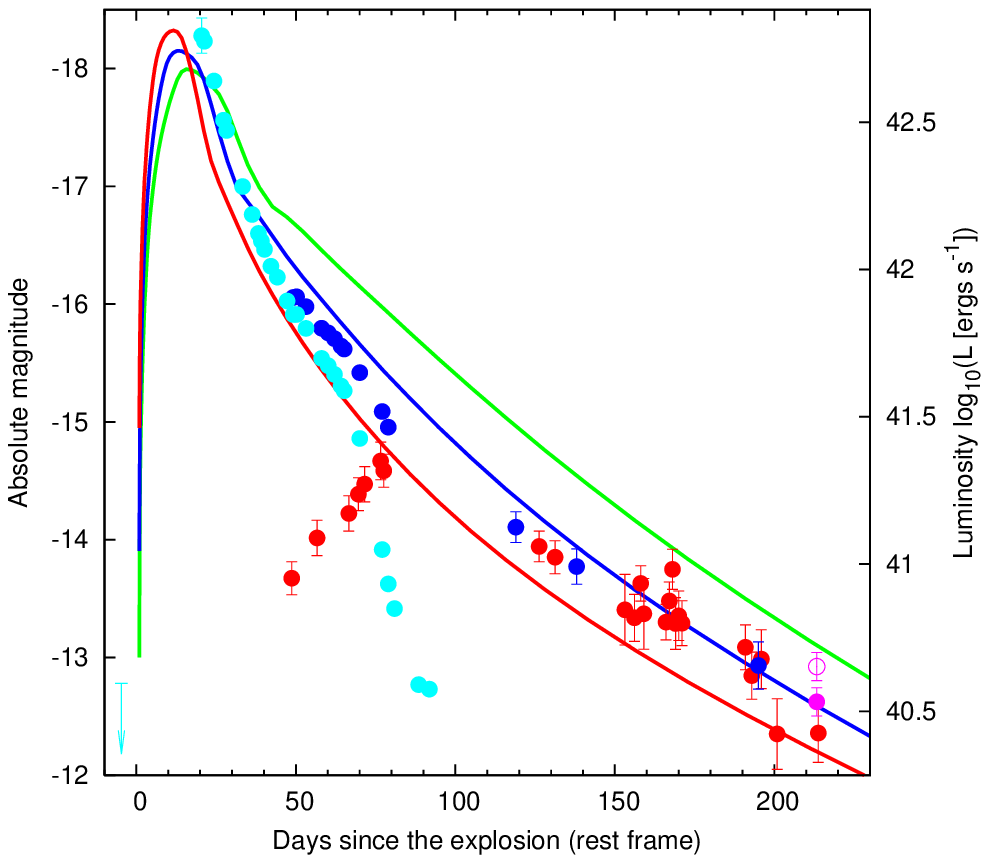}
\caption{Comparison between the synthetic LCs for the models with
 $E_{51}=5$ and $\Mej=5.1\Msun$ ({\it green line}), $E_{51}=10$ and
 $\Mej=4.9\Msun$ ({\it blue line}), and $E_{51}=20$ and $\Mej=4.6\Msun$
 ({\it red line}) and the LCs of SN~2006jc ($L_{\rm UV}+L_{\rm opt}$:
 {\it cyan filled circles}, \citealt{imm07,kaw07a,kaw07b,anu07}; 
 $L_{\rm IR,est}(\nu<3\times10^{14}{\rm Hz})$: {\it red filled circles},
 \citealt{ark06,dic07}; $L_{\rm bol}$: {\it blue filled circles},
 \citealt{imm07,kaw07a,kaw07b,anu07,ark06,dic07}; $L_{\rm IR,hot}(\nu<3\times10^{14}{\rm Hz})$:
 {\it magenta filled circle}, \citealt{sak07}; $L_{\rm IR}(\nu<3\times10^{14}{\rm Hz})$: {\it
 magenta open circle}, \citealt{sak07}). The
 luminosities denoted by the circles without errorbars are considerably
 contributed by the UV luminosity estimated as $L_{\rm UV}=0.8L_{\rm opt}$.
\label{fig:LC}}
\end{figure}

The peak width ($\tau$) of the SN LC depends on the ejected mass $\Mej$,
explosion energy $E$, opacity $\kappa$, density structure, and \Nifs\
distribution, as $\tau\propto A\kappa^{1/2} \Mej^{3/4} E^{-1/4}$
\citep{arn82}, where $A$ represents the effects of the density structure
and the \Nifs\ distribution. Here, we assume for sake of simplicity a uniform
mixing of \Nifs\ in the SN ejecta. Also, the density structures after the
SN explosions with various $E$ are analogous. Thus, the dependence on
$A$ is negligible and we investigate the LC properties depending on
$\kappa$, $\Mej$ and $E$. The synthetic LCs obtained for the models with
($\Mej/\Msun$, $E_{51}$) $=$ (5.1, 5), (4.9, 10), and (4.6, 20) are
shown in Figure~\ref{fig:LC}. Figure~\ref{fig:LC} also shows 
the multicolor and bolometric LCs of SN~2006jc.

\subsection{Comparison with Observations}

The period of SN~2006jc is divided into four epochs depending on the
available observations: (1) UV and optical photometries at $t<50$ days,
(2) optical and NIR photometries at $t\sim50-80$ days, (3) optical
photometry at $t\sim80-120$ days, and (4) optical, NIR, and MIR
photometries and NIR spectroscopy at $t>120$ days.

(1) At $t<50$ days, the IR contributions to the bolometric luminosity
may well be small because the IR contribution is only $\sim10\%$ at
$t\sim50$ days. Thus, the peak bolometric luminosity derived from the UV
and optical fluxes is reliable (\S~\ref{sec:OPTest}). If the bolometric LC peaked at the
discovery, the peak luminosity is reproduced by the \Nifs-\Cofs\ decay of
$\Mni=0.22\Msun$. 
The rapid decline after the peak prefers such high-energy models as ($\Mej/\Msun$,
$E_{51}$) = (4.9,~10) and (4.6,~20). 

(2) At $t\sim50-80$ days, the IR contribution to the bolometric
luminosity increases from $\sim10\%$ at $t=49$ days to $\sim70\%$ at
$t=79$ days. The contribution of $L_{{\rm IR,est}}(\nu<1.3\times10^{14}{\rm Hz})$ 
to $L_{{\rm IR,est}}(\nu<3\times10^{14}{\rm Hz})$ changes from $\sim30\%$ at $t=49$ days
to $\sim40\%$ at $t=79$ days. At this epoch, the 
optical and IR emissions contribute to the bolometric
luminosities. Combining the IR brightening and the optical
decline, the bolometric LC including $L_{\rm UV}$ ($=0.8L_{\rm opt}$)
declines slowly. Such a slow decline is consistent with the models
of ($\Mej/\Msun$, $E_{51}$) = (4.9,~10) and (4.6,~20).

(3) At $t\sim80-120$ days, NIR photometries are
not available. The decline of the optical luminosity at
this epoch is more rapid than at $t<80$ days. Such a rapid decline of the
optical LC can not be reproduced by the \Nifs-\Cofs\ decay. However, the
bolometric LC may well decline more slowly than the optical LC because
the IR emission dominates in the bolometric luminosity. 

(4) At $t>120$ days, NIR photometries are
available continuously and optical photometries are
available at $t=120,~140,~{\rm and}~195$ days \citep{kaw07b}. The contribution of
the optical emission to the bolometric luminosities is negligible
($\sim3\%$ at $t=120,~140,~{\rm and}~195$ days). 
At this epoch, the contribution of $L_{{\rm IR,est}}(\nu<1.3\times10^{14}{\rm Hz})$ 
to the total luminosity increases from $\sim60\%$ at $t=127$ days to $\sim80\%$ at $t=215$
days. The estimated IR luminosity is consistent with the
luminosity emitted from the hot carbon dust at $t=215$ days (\S~\ref{sec:IRest}). 
Since the dust temperature decreases with time, the ratio of the
MIR luminosities to the NIR luminosities becomes
larger with time. Therefore, the amorphous carbon emission model
reasonably estimates the IR luminosity due to the hot carbon dust at
$t\lsim215$ days. The model with $\Mej=4.9\Msun$, $E_{51}=10$, and
$\Mni=0.22\Msun$ reproduces well the LC decline at $t>120$ days and the
IR luminosities due to the hot carbon dust at $t=215$
days\footnote{After submission of this paper,
the IR observation at $t=425$ days was presented by \cite{mat08}. They
estimated the total luminosity and the dust temperature as
$L\sim1.2\times10^{40}$\ergs\ and $T=520$~K, respectively. Our model with $\Mej=4.9\Msun$,
$E_{51}=10$, and $\Mni=0.22\Msun$ predicted the luminosity 
of $L_{\rm bol}\sim3\times10^{39}$\ergs and the dust temperature of
$T\sim200$~K \citep{noz07} at $t\sim430$ days, which are lower than the
observations. This suggests that the observed IR emissions at $t=425$
days may originate not only from the newly-formed dust in the SN ejecta 
heated by the \Nifs-\Cofs\ decay but also from the light echo of the CSM dust.}.
Therefore, we conclude that the hypernova-like SN explosion model with
$\Mej=4.9\Msun$, $E_{51}=10$, and $\Mni=0.22\Msun$ is the most
preferable model among the exploded models of a WCO Wolf-Rayet star with
$\Mpre=6.9\Msun$.

\section{Interaction with Circumstellar Matter}
\label{sec:CSM}

X-rays from SN~2006jc were detected by the {\sl Swift} and {\sl
Chandra} satellites \citep{imm07}.  The X-ray detection indicates
that the expanding SN ejecta collides with the CSM.

We calculate X-ray emission from the ejecta-CSM interaction for the
SN model with ($\Mej/\Msun$, $E_{51}$) $=$ (4.9, 10), and estimate the
CSM density structure on the basis of a comparison with the observed X-ray LC (\eg
\citealt{suz95}).  The observed X-ray luminosities estimated with the
distance of $24$ Mpc in \cite{imm07} are scaled using $25.8$ Mpc.

We adopt a CSM density profile characterized by a power-law of 
 $\rho = \rho_0 (r/r_0)^{-n}$ and assume that the interaction starts at a distance
$r=3\times10^{10}$ cm.  The parameters $\rho_0$, $r_0$, and $n$ are
determined so that the ejecta-CSM interaction reproduces the observed
X-ray LC.

\begin{figure}
\plotone{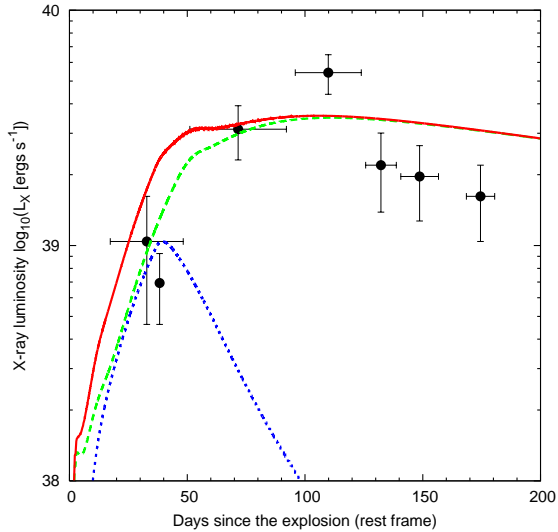}
\caption{Comparison between the synthetic X-ray LCs contributed from the
 total emission ({\it solid line}), the SN ejecta ({\it dashed line}),
 and the CSM ({\it dotted line}) and the X-ray LC of SN~2006jc ({\it
 circles}, \citealt{imm07}).
 \label{fig:Lx}}
\end{figure}

The interaction generates reverse and forward shock waves in the
SN ejecta and CSM, respectively. 
Both regions are
heated by the shock waves and emit X-rays. In such a compact star,
because the density in the shocked SN ejecta is higher than that in
the shocked CSM, the emitted X-rays from the shocked SN ejecta are more
luminous than those from the shocked CSM (Fig.~\ref{fig:Lx}).

Figure~\ref{fig:Lx} shows the synthesized X-ray LC for 
$\rho_0=2.75\times10^{-19}$ g cm$^{-3}$ and $n = 0$ for 
$r < 2.2\times10^{16}$
cm and $n = 6$ for $r > 2.2\times10^{16}$ cm, i.e., for a flat
(inside) and steep (outside) CSM density profile of 
$\rho =2.75\times10^{-19}$ g cm$^{-3}$ for $r < 2.2\times10^{16}$ cm and
$2.75\times10^{-19} (r / 2.2\times10^{16} {\rm cm})^{-6}$ g cm$^{-3}$
for $r > 2.2\times10^{16}$ cm.  The total mass of the CSM is
$1.2\times10^{-2} M_\odot$ to reproduce the peak of the
observed X-ray LC and the subsequent decline.

\begin{figure}
\plotone{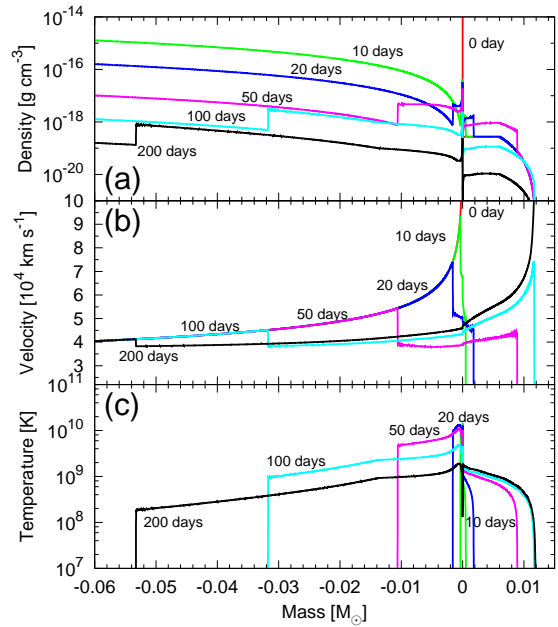}
\caption{(a) Density, (b) velocity, and (c) temperature structures of
 the SN ejecta and the CSM at $t=0$ day ({\it red}), $10$ days ({\it
 green}), $20$ days ({\it blue}), $50$ days ({\it magenta}), $100$ days
 ({\it cyan}), and $200$ days ({\it black}). The coordinate is the
 Lagrangian mass with the contact discontinuity between the ejecta ({\it
 left}) and the CSM ({\it right}). 
\label{fig:CSM}}
\end{figure}

The density, velocity, and temperature structures and their evolutions
are shown in Figures~\ref{fig:CSM}abc. The velocity of the reverse shock is
$v\sim3.8\times10^4$ \kms. The reverse shock reaches
$\sim5.3\times10^{-2} \Msun$ from the outer edge of the SN ejecta at
$t=200$ days and heats up the swept-up SN ejecta. The temperature behind the
reverse shock is higher than $10^8$ K where dust cannot newly form and the
dust formed in the SN ejecta is destroyed
\citep{noz07}. Our calculation does not show the formation of a cooling
shell. This is because the CSM
interaction is so weak to emit X-ray of $\sim3\times10^{39}$
\ergs. If the bolometric luminosity is powered by the CSM
interaction, \ie if the CSM interaction emits as high luminosity as 
$\sim8\times10^{42}$ \ergs, the cooling shell might form and thus the
dust formation might be possible. Further detailed studies, however, are required
to confirm the dust formation behind the reverse shock.

Such a flat density profile of the inner CSM implies that the stellar wind was not
steady because the steady wind should form the CSM of 
$\rho \propto r^{-2}$.  This circumstellar environment might have been formed by a
variable mass-loss rate $\dot{M}$ and/or a variable wind velocity
$v_{\rm w}$. For example, assuming that the stellar wind blew with a
constant $v_{\rm w} = 3,500$ km s$^{-1}$ for two years before the
explosion, the mass-loss rate must have changed from $1\times10^{-2}$ to
$2\times10^{-14}$ $M_\odot$ yr$^{-1}$ in two years.  
Such a drastic change of the mass-loss rate and/or the wind velocity is consistent
with the fact that the progenitor of SN~2006jc was surrounded by the
matter ejected by the LBV-like event two years before the explosion.

\section{Conclusions}
\label{sec:conclude}

We present a theoretical model for SN~2006jc whose properties are
summarized as follows.

(1) {\bf WCO progenitor and Dust Formation}: The progenitor is a WCO
Wolf-Rayet star whose total mass has been reduced from $\Mms=40\Msun$ 
to as small as $\Mpre=6.9\Msun$. The WCO
star model has a thick C-rich envelope and CSM. This is consistent
with the formation of amorphous carbon grains in the SN ejecta and
the CSM suggested by {\sl AKARI} observations
\citep{sak07}. \cite{noz07} have calculated dust formation
in the WCO star explosion model and shown that carbon dust is formed in the C-rich layer
at $t\sim50$ days. This is much earlier than the dust formation after
$t\sim1$ yr in Type II SNe (SNe~II), because of the much smaller $\Mej$
in the WCO star than SNe~II. 
According to the models in \cite{lc06}, the stars with
$\Mms>40\Msun$ typically become WCO stars to form a thick C-rich layer
and CSM. This limiting mass, however, is still uncertain and strongly depends
on many details of the stellar evolution.  
The early dust
formation in the SN ejecta and the CSM suggests that the progenitor of
SN~2006jc is a massive star becoming a WCO Wolf-Rayet star.

(2) {\bf Explosion and Bolometric Light Curve}: The multicolor LCs
of SN~2006jc show peculiar evolutions, \eg a rapid decline of the optical
LC and brightening of the IR LC. These can be interpreted as an ongoing
dust formation. Assuming the absorbed optical light is re-emitted in the IR
band, the bolometric LC is constructed as a summation of 
$L_{\rm UV}$ ($=0.8L_{\rm opt}$), $L_{\rm opt}$, and $L_{\rm IR}$. By
calculating the hydrodynamics, nucleosynthesis, and the bolometric LC
for the SN explosion with the various explosion energies, $E_{51}=1$, 5,
10, and 20, we find that the hypernova-like SN
explosion model with $\Mej=4.9\Msun$, $E_{51}=10$, and $\Mni=0.22\Msun$
best reproduces the bolometric LC of SN~2006jc with the
radioactive decays. Also, the temperature evolution of the
carbon dust heated by the \Nifs-\Cofs\ decay reasonably well explains the
IR observations for $t\sim50-220$ days \citep{noz07}.

(3) {\bf CSM Interaction and X-ray Light Curve}: Applying the model with
$\Mej=4.9\Msun$ and $E_{51}=10$, we calculate
the ejecta-CSM interaction and the resultant X-ray LC. We derive the 
CSM density structure to reproduce the X-ray LC of SN~2006jc as 
$\rho =2.75\times10^{-19}$ g cm$^{-3}$ for $r < 2.2\times10^{16}$ cm and
$2.75\times10^{-19} (r / 2.2\times10^{16} {\rm cm})^{-6}$ g cm$^{-3}$
for $r > 2.2\times10^{16}$ cm.
The flat density distribution in the inner CSM indicates a drastic
change of the mass-loss
rate and/or the wind velocity that is consistent with the LBV-like event
two years before the explosion.

\vspace{1.5cm}

\section{Discussion}
\label{sec:discuss}

{\bf LBV connection}: 
Our model does not take into account the LBV-like
event that occurred two years before the explosion.
The first reason is that the mechanism of the
outburst is still unclear. The second reason is that, at least in the
framework of the current
understanding of standard stellar evolution, the envelope of
a massive star practically freezes out after core He
exhaustion (i.e., about 10000 years before the explosion) 
due to the more rapid evolution of the core than the
envelope. In addition, it is interesting to note that, and this
is a confirmation of the theoretical expectation,
there is no observational evidence that any Wolf-Rayet star
has ever undergone such a luminous outburst \citep{hum99}. 
Hence, there is no specific reason 
to associate the occurrence of a LBV-like outburst
to the presupernova evolution.
Future studies on the mechanism of the
outburst are required to firmly conclude the origin of the LBV-like
outburst. It would be possible that a possible binary companion star
could undergo the LBV-like outburst.

{\bf Fallback}: According to our hydrodynamics and nucleosynthesis
calculations, in the spherically symmetric models with $E_{51}\geq5$,
the fallback does not take place and thus the amount of synthesized \Nifs\ is
much larger than $\Mni=0.22\Msun$ which is required to power the LC
of SN~2006jc (\S~\ref{sec:56ni}). In the aspherical explosions, however,
the fallback takes place even for $E_{51}\geq5$. In this paper, we assume the
fallback even for the models with $E_{51}\geq5$ and derive the amount
of fallback to yield the appropriate amount of \Nifs. 
To justify the above assumption, we calculate an aspherical explosion
induced by a jet with an opening angle of $\theta=45^\circ$ and an
energy deposition rate of $\dot{E}=3\times10^{52}~{\rm ergs~s^{-1}}$
\citep{tom07a,tom07b}.
The jet-induced model realizes an explosion with $\Mej\sim4.9\Msun$,
$E_{51}\sim10$, and $\Mni\sim0.22\Msun$ that is consistent with the
adopted model. We note that an aspherical
radiative transfer calculation is required to confirm that the
jet-induced explosion model can reproduce the LC of SN~2006jc. 

{\bf Light curve models}: The model with $\Mej=4.9\Msun$, $E_{51}=10$, and $\Mni=0.22\Msun$ is not
an unique model to reproduce the bolometric LC of SN~2006jc. 
In the case of usual SNe, the velocities of the absorption lines can
disentangle the degeneracy of $\Mej$ and $E$ by means of the comparison
with the photospheric velocities (\eg \citealt{tan08}). However, the
spectra of SN~2006jc are 
dominated by He emission lines and the nature is unclear. Thus we cannot
fully resolve the degeneracy. Since the LC shape is proportional to
$\Mej^{3/4} E^{-1/4}$, the model with a larger $\Mej$ requires a higher $E$.
An explosion of the progenitor star with a larger $\Mpre$ 
reproduces the LC of SN~2006jc with a higher $E$ and the X-ray LC with a
lower CSM density. On the
other hand, an explosion of the progenitor star with a smaller $\Mpre$
reproduces the LC with a lower $E$, \eg an explosion
with $\Mej=1.5\Msun$ and $E_{51}=1$ can explain the LC shape of
SN~2006jc. However, such low $E$ explosions suppress explosive
nucleosynthesis and enhance the fallback. As a result, the \Nifs\
production is reduced for a small $\Mpre$.
If the LC of SN~2006jc is powered by the \Nifs-\Cofs\ decay, the bright
peak indicates a larger amount of \Nifs\ production ($\sim0.22\Msun$) than
a normal SN [$\Mni\sim0.07\Msun$, \eg SN~1987A, \citealt{bli00}]. Therefore,
SN~2006jc is likely a more energetic explosion than a normal SN with
$E_{51}\sim1$.

{\bf Dust formation}: We assume that the energy source of the LC of SN~2006jc is the
\Nifs-\Cofs\ decay. This consistently explains the formation of carbon
dust at the early epoch ($t\sim50$days) and the dust 
temperature at $t\sim200$days \citep{noz07}. In this scenario, however,
the origin of the bright blue continuum remains an unsolved
problem (\eg \citealt{pas07,smi07,imm07}). Such a spectrum might be
explained by the ejecta-CSM interaction. In this scenario, however, the
fine tunings are required to reproduce the bolometric LC; most of the
X-rays are absorbed and converted to the optical luminosity, which only a small
fraction of the X-rays are emitted with changing the fraction
from $10^{-3}$ at $t\sim30$ days to $0.1$ at $t\sim180$ days.
Moreover, the formation of carbon dust with two temperatures would not
be explained. Since both scenarios are inconclusive so far, further
investigations may give important implications on the emission mechanism
of SN~2006jc. 

\acknowledgements

We would like to thank A.~Arkharov, N.~Efimova, A.~Di~ Paola,
C.~Corsi, E.~Di~Carlo, and M.~Dolci for providing us the data of the JHK-band
photometries. We also would like to
thank S. Immler for the preprint on the Swift observations.
M.L. thanks the support from the 21st Century COE Program (QUEST) 
of JSPS (Japan Society for the Promotion of Science) for his
stay in the University of Tokyo and the hospitality of the Department of
Astronomy. G.C.A., D.K.S., and T.P.P. are supported by the 
JSPS - INSA (Indian National Science Academy) exchange programme.
N.T. and M.T. are supported through the JSPS Research Fellowship for Young Scientists.
This work has been supported in part by World Premier International
Research Center Initiative (WPI Initiative),
MEXT, Japan, and by the Grant-in-Aid for Scientific
Research of the JSPS (10041110, 10304014, 11740120,
12640233, 14047206, 14253001, 14540223, 16740106,
18104003, 18540231, 20540226) and MEXT (19047004,
20040004, 20041005, 07CE2002).

\end{document}